\documentclass[journal,article,accept,moreauthors,pdftex,12pt,a4paper,entropy]{mdpi}
\usepackage{}
\lastpage{0}
\doinum{x/entropy}
\pubvolume{x}
\pubyear{2016}
\history{Received: xx / Accepted: xx / Published: xx}

\Title{Universal entropy relations: entropy formulae and entropy bound}

\Author{Hang Liu $^{b}$, Xin-he Meng $^{b}$, Wei Xu $^{a}$* and Bin Zhu$^{c}$}

\address{%
$^{a}$ MOE Key Laboratory of Fundamental Physical Quantities Measurements, School of Physics, Huazhong University of Science and Technology, 1037 Luoyu Road, Wuhan 430074, China\\
$^{b}$ School of Physics, Nankai University, Tianjin 300071, China\\
E-mails: xhm@nankai.edu.cn\\
$^{c}$ Institute of Physics Chinese Academy of sciences, Beijing 100190, China\\
E-mails: binzhu7@gmail.com}

\corres{E-mail: xuweifuture@gmail.com.}

\abstract{
We survey the applications of universal entropy relations in black holes with multi-horizons. In sharp distinction to conventional entropy product, the entropy relationship here not only improve our understanding of black hole entropy but was introduced as an elegant technique trick for handling various entropy bounds and sum. Despite the primarily technique role, entropy relations have provided considerable insight into several different types of gravity, including massive gravity, Einstein-Dilaton gravity and Horava-Lifshitz gravity. We present and discuss the results for each one.
}



\begin{document}


\section{Introduction}

Ideas from the theory of black hole thermodynamics have been of great importance in understanding gravity ever since Bardeen, Carter and Hawking showed how the two subjects can be connected \cite{Bardeen:1973gs}. Almost every major development in thermodynamics leads to a parallel progress in the study of black holes in the last fifty years. For example, the black temperature is thought of being its surface gravity. However the subject is still not fully understood especially when the quantum effects are considered.

Quantum effects induces great difficulty for the black hole to find its microscopic origin of entropy. Although there are some significant progress in calculating the black hole entropy, such as Kerr/CFT correspondence, the calculation is limited to the special classes of black holes including BPS black holes \cite{Guica:2008mu} (One can see the recent review paper \cite{Compere:2012jk} for the generalizations and extensions of the Kerr/CFT correspondence). As a consequence an phenomenological approach called universal area product is proposed to deal with general black holes, which make it necessary to include the effect of multi-horizons \cite{Cvetic:2010mn}. There already exist excellent papers on universal entropy relations (UER) and it may therefore be important to say a few words on the new character in which the paper should be placed. We have put the emphasis on calculating entropy product which does not depend on the mass. For completeness, the entropy sum in AdS spacetime is mentioned. Here we found the entropy bounds are consistent with Penrose inequality and related closely to the cosmological radius. Having this in mind, we can introduce a unified description of non-minimal gravity in terms of UER. The simplest examples are massive gravity, Einstein-Dilaton gravity and Horava-Lifshitz gravity, whose results are given explicitly in this paper.

The paper is structures as follows: in section 2, we recapitulate universal entropy product and some failed examples; in section 3, the UER as well as its extensions are given; in section 4, we deal with the entropy bounds; the last three sections are devoted to study the UER in various non-minimal gravity models.

\section{Universal entropy (area) product}

We recapitulate the universal entropy product (UEP) proposed by Cevtic et al \cite{Cvetic:2010mn}. Moreover we mention that the produce is only the function of black hole hires which are charges, angular momenta and cosmological constant respectively. A transition to more complicated situations is easy when we start with UEP. Consider a multi-charged black hole \cite{Cvetic:1996kv,Cvetic:1996xz}, the entropy of the outer horizon can be written as \cite{Cvetic:1996kv}
\begin{align}
S_{+}=2\pi(\sqrt{N_L}+\sqrt{N_R}),
\end{align}
while the inner one takes the form \cite{Larsen:1997ge,Cvetic:1997uw,Cvetic:1997xv},
\begin{align}
S_{-}=2\pi(\sqrt{N_L}-\sqrt{N_R}),
\end{align}
where $N_L$ and $N_R$ are identified as the number of left-moving and right-moving modes of conformal field theory. Then the entropy product can read easily,
\begin{align}
S_{p}=S_{+}S_{-}=4\pi^2(N_L-N_R).
\end{align}
In terms of the duality conjecture, $S_p$ must be function of charges $Q_i$ and angular momentum $J$ \cite{Cvetic:2010mn},
\begin{align}
S_p&=4\pi^2(\prod_{i=1}^{4} Q_i+J^2),\label{eqn:4d}\\
S_p&=4\pi^2(\prod_{i=1}^{3} Q_i+J_a J_b),\label{eqn:5d}
\end{align}
where equation $\ref{eqn:4d}$ and equation $\ref{eqn:5d}$ correspond to the $4$-dimension and $5$-dimensional black hole solutions respectively. Furthermore $4$-dimensional black hole solution is described by four charges $Q_i$ and one angular momentum $J$, while the $5$-dimensional solution is described by three charges and two angular momentum $J_a,J_b$. Such a multi-horizon solution is too complicated to calculate while the area product is still universal for rotating black holes in gauge supergravities \cite{Cvetic:2010mn,Toldo:2012ec,Cvetic:2013eda,Lu:2013ura,Chow:2013tia}. The study can also be generalized to charged rotating black hole in various gravity models (including Einstein gravity \cite{Castro:2012av,Visser:2012zi,Chen:2012mh,Castro:2013kea,Abdolrahimi:2013cza,Xu:2014qaa}, $f(R)$ gravity\cite{Castro:2013pqa,Cvetic:2013eda,Xu:2014qaa}, higher curvature gravity \cite{Castro:2013pqa,Cvetic:2013eda,Xu:2014qaa} and other models with matter fields \cite{Castro:2013pqa,Lu:2013eoa,Anacleto:2013esa,Xu:2014qaa,Page:2015gia,Pradhan:2015yea,Pradhan:2015fha,Majeed:2015eia,Pradhan:2015ela,Mandal:2015ozx} ), and the result shows that $S_p$ is also independent of mass of black hole.

\subsection*{Failed examples}

If the area product is universal for general black holes, people can obtain more information of entropy from the dual conformal field theory. However property of mass independence is not always available. Consider a black hole with vanishing charges and angular momenta, one can find that the above universal area product is zero, thus leads to vanishing area of horizons corresponding to the vacuum. This unreasonable result indicates that the universal area product breaks for black holes without charges and angular momentum. Take Schwarzschild-de Sitter black hole as an example \cite{Visser:2012wu},
\begin{align}
ds^2&=-f(r)dt^2+\frac{dr^2}{f(r)}+r^2(d\theta^2+\sin^2\theta d\psi^2)\nonumber\\
f(r)&=1-\frac{2M}{r}-\frac{\Lambda r^2}{3},
\end{align}
where $M$ is the mass of black hole and $\Lambda=1\l^2$ is positive cosmological constant. The solution is associated with two different types of horizons, i.e. even horizon $r_E$ and cosmological horizon $r_C$,
\begin{align}
r_E&=2l\sin\left(\frac{1}{3}\left(\frac{3M}{l}\right)\right),\nonumber\\
r_C&=2l\sin\left(\frac{1}{3}\left(\frac{3M}{l}\right)+\frac{2\pi}{3}\right).
\end{align}
The third root of metric function $r_V$ is negative and called virtual horizon which is not physical. Notice that the product of areas of physical horizons,
\begin{align}
A_E A_C=64\pi^2 l^4\left(\frac{1}{2}-\cos\left(\frac{2\pi}{3}+\frac{2\pi}{3}\arcsin\left(\frac{3M}{l}\right)\right)\right)
\end{align}
has no nice quantization features. Even including the effect of the third unphysical virtual horizon one can find,
\begin{align}
A_E A_C A_V=2304\pi^3 M^2 l^4,
\end{align}
which is not mass independent. Therefore the UER can not provide any quantization information. For Schwarzschild-AdS black hole, the mass independence is still lost. The absence of charge and angular momenta seems to break the universality of multi-horizons black holes \cite{Castro:2013pqa,Xu:2014qaa,Faraoni:2012je}. Moreover, we find even including the effect of charge and rotation, e.g. the Reissner-Nordstorm black holes \cite{Visser:2012wu} and Kerr-Newman-Taub-NUT black holes \cite{Pradhan:2013hqa,Pradhan:2013xha}, the UER still can not work. In this sense we are left with the charged rotating black holes with multi-horizons, which are the only candidate of mass independence.

\section{Further extensions of UER}\label{Section3}

Follow the spirit of the area product, more area relations including quantization features are present. These interesting relations are composed of partial entropy product and entropy summation, both of which have the nice property of mass independence. The partial entropy product is introduced along with the mass dependent area product in the last section. It behaves as \cite{Xu:2014qaa,Wang:2013nvz}
\begin{align}
\sum (S_i S_j)^{\frac{1}{d-2}}=-\frac{k(d-1)(d-2)\pi^{(d-1)/(d-2)}}{2\Lambda(2\Gamma(d-2/2))}
\end{align}

We can find that it is independent of mass, charge and angular momentum. This opens a new window on calculation. However the partial entropy product can not be generalized to modified gravity case. Entropy sum also has universality for multi-horizon black holes, even there is no charge and angular momenta,
\begin{align}
S_{+}+S_{-}=4\pi\sqrt{N_L},
\end{align}
which is only the function of left moving modes. Actually entropy sum can be expressed in terms of $\Lambda$. It is easy to find in $4$-dimensional spacetime \cite{Wang:2013smb},
\begin{align}
\sum_i S_i=\frac{6\pi}{\Lambda}.
\end{align}
No matter what spacetime we study, the results hold for true in Einstein gravity. In non-minimal gravity models, the relation gets slight modification. For example in $f(R)$ gravity and Einstein-scalar gravity we have \cite{Wang:2013smb,Xu:2013zpa},
\begin{align}
\sum S_i&=\frac{6\pi}{\Lambda_f}(1+f^{\prime}(R_0)),\nonumber\\
\sum S_i&=\frac{6\pi}+\frac{\pi}{6\alpha}.
\end{align}
The universality of entropy sum are also checked in Einstein-Weyl gravity and gauged supergravity \cite{Wang:2013smb,Xu:2013zpa}. A strict proof for the universality of entropy sum in general dimensions is given in \cite{Du:2014kpa}, where the entropy sum only depends on the topology of the black hole. However, there are also some counterexamples for the universal entropy sum \cite{Xu:2013zpa}. It is easy to find that the entropy sum diverges at the zero cosmological constant limit, hence the universality might fail in asymptotically flat spacetime. Besides one can formulate a general criterion for the validity of the universality \cite{Zhang:2014ada}:
\begin{enumerate}
\item Consider a $d$-dimensional spacetime with spherical symmetry, whose metric function $f(r)$ is a Laurent series within the range $m>d-2$ and $n>4-d$. Here $m$ is the lowest power and $n$ is the highest power.
\item Generalize to rotating black hole, i.e. Myers-Perry black hole. The entropy product and entropy sum are both mass independent for $d>4$.
\end{enumerate}

\section{Entropy bounds}

Though more UER are introduced for black holes, it is still ambiguous to connect the microscopic origin with the black hole entropy directly. Motivated by the UER development, a great improvement in understanding entropy bounds is induced which are related closely to the holography principle \cite{Bousso:1999xy}. Here we give some entropy bounds from entropy relations which are consistent with Penrose inequality \cite{Mars:2009cj}. In particular the entropy bound in AdS spacetime becomes a geometric inequality which is the function of cosmological radius.

\subsection{Asymptotically flat black hole}

Firstly let us briefly review the entropy bounds of four dimensional Kerr black holes with vanishing cosmological constant \cite{Xu:2015mna},
\begin{align*}
  \mathrm{d}s^2&=-\frac{\Delta-a^2\sin^2\theta}{\Sigma}\mathrm{d}t^2-2a\sin^2\theta\left(\frac{r^2+a^2-\Delta}{\Sigma}\right)\mathrm{d} t\mathrm{d}\theta\\
  &+\left(\frac{(r^2+a^2)^2-\Delta\,a^2\sin^2\theta}{\Sigma}\right)\sin^2\theta\mathrm{d}\phi^2+\frac{\Sigma}{\Delta}\mathrm{d}r^2+\Sigma\mathrm{d}\theta^2,\\
 &\Sigma=r^2+a^2\cos^2\theta,\quad\Delta=r^2-2Mr+a^2,
\end{align*}
from the metric function $\Delta$, one can find two horizons: the event horizon $r_{E}=M+\sqrt{M^2-a^2}$ and Cauchy horizon $r_{C}=M-\sqrt{M^2-a^2}$. The area entropy $S_{i}=\frac{A_{i}}{4}=\pi(r_{i}^2+a^2)$ leads to
the mass-independent entropy product
\begin{align}
S_{E}S_{C}=4\pi^2 J^2 ,
\end{align}
and the mass-dependent entropy sum (where $J=Ma$)
\begin{align}
S_{E}+S_{C}=4\pi M^2,
\end{align}
which is consistent with the result of rotating black holes in asymptotically flat spacetime.

In order to avoid naked singularities, we introduce the famous Kerr bound
$  M\geq\,a, $ i.e. $M^2\geq\,J.$
Then one can easy derive the entropy bounds for event horizon and Cauchy horizon
\begin{align}
S_{E}\in\bigg[2\pi\,M^2,4\pi\,M^2\bigg],\quad\,S_{C}\in\bigg[0,2\pi\,J\bigg],
\end{align}
and the area bounds of event horizon and Cauchy horizon
\begin{align}
\sqrt{\frac{A_{E}}{16\pi}}\in\bigg[\frac{M}{\sqrt{2}},M\bigg],\quad\,\sqrt{\frac{A_{C}}{16\pi}}\in\bigg[0,\sqrt{\frac{J}{2}}\bigg],
\end{align}
where the upper bound of area for event horizon is actually the first geometrical inequality of black holes, i.e. the Penrose inequality \cite{Mars:2009cj}.

These geometrical area (entropy) bounds are also studied in Kerr black hole family \cite{Xu:2015mna}, including the  Kerr-Newman black hole and  Kerr-Taub-NUT black hole. It is found that
the electric charge $Q$ diminishes the geometrical area (entropy) bound for event horizon, while it enlarges that for Cauchy horizon; the angular momentum $J$ enlarges them
for Cauchy horizon, while it does nothing with that for event horizon; the NUT charge always enlarges them for both event horizon and Cauchy horizon.

\subsection{Asymptotically AdS black hole}

Then we show the entropy bounds in AdS spacetime. Note the cosmological constant is interpreted
as thermodynamic pressure, which should be treated as a thermodynamic variable in the thermodynamics phase space (see, e.g. \cite{Kastor:2009wy,Dolan:2011xt,Cvetic:2010jb,Dolan:2013ft,Altamirano:2014tva}). For this case, the entropy bound becomes a geometrical inequality which is related closely to the cosmological radius.

We introduce the entropy bounds of four dimensional Schwarzschild-dS black hole here \cite{Xu:2015eia}. The area entropy $S_{i}=\frac{A_{i}}{4}=\pi r_i^2,\quad\, (i=E,C,V)$ lead to the mass-dependence entropy product
\begin{align}
  S_{E}S_{C}S_{N}=36\pi^3M^2\ell^4;
\end{align}
the mass-independence entropy sum
\begin{align}
  S_{E}+S_{C}+S_{N}=6 \pi \ell^2;
\end{align}
and the mass-independence ``part'' entropy product
\begin{align}
  S_{E}S_{C}+S_{E}S_{N}+S_{C}S_{N}=9\pi^2\ell^4.
\end{align}
Note the final one is consistent with another mass-independent UER \cite{Visser:2012wu,Xu:2014qaa}
\begin{align}
  S_{E}+S_{C}+\sqrt{S_{E}S_{C}}=3\pi \ell^2,\label{SdSTwo}
\end{align}
where only the effect of two physical horizons is included.

For the physical acceptable black holes, we introduce the mass bound
$\frac{3 M}{\ell}\leq1$.
Then after a little calculation, we can obtain the entropy bounds of all horizons
\begin{align}
\,S_{E}\in\bigg[0,\pi\ell^2\bigg],\quad\,S_{C}\in\bigg[\pi\ell^2,3\pi\ell^2\bigg],\quad\,S_{N}\in\bigg[3\pi\ell^2,4\pi\ell^2\bigg],
\end{align}
together with the area bound
\begin{align}
\sqrt{\frac{\,A_{E}}{16\pi}}\in\bigg[0,\frac{\ell}{2}\bigg],\quad\,\sqrt{\frac{\,A_{C}}{16\pi}}\in\bigg[\frac{\ell}{2},\sqrt{\frac{3}{4}}\ell\bigg],\quad\,\sqrt{\frac{\,A_{N}}{16\pi}}\in\bigg[\sqrt{\frac{3}{4}}\ell,\ell\bigg].
\end{align}
which are all geometrical bounds, as parameter $\ell$ is actually the cosmological radius.

These geometrical area (entropy) bounds are also studied in Gauss-Bonnet gravity \cite{Xu:2015eia}, where the Gauss-Bonnet coupling constant should also be  treated as a thermodynamical variable (see, e.g. \cite{Kastor:2010gq,Cai:2013qga,Xu:2013zea,Xu:2014tja,Xu:2014kwa,Altamirano:2014tva}). Meanwhile, area bounds are modified by the constants of theory.

In the following words of this paper, we generalized the above results to modified gravity to improve
the discussion of general black holes with multi-horizons. Note that the coupling constants of theory may also be  treated as a thermodynamical variable. We expand the discussion about entropy relations in massive gravity, Einstein-dilaton gravity and Horava-Lifshitz gravity. We also obtain thermodynamic laws and new entropy bounds in these modified gravity models. These results provide further insights into the crucial role played by the entropy relations of multi-horizons in black hole thermodynamics.

\section{Entropy relations and bounds in massive gravity}

In this section, we study entropy relations and bounds in massive gravity with the following action in $n+2$ dimensional spacetime \cite{Vegh:2013sk}
\begin{align*}
S=\frac{1}{16\pi}\int \mathrm{d^{n+2}}x\sqrt{-g}\left[
R-2\Lambda+m^2\sum_{i=1}^4c_i U_i(g,f)\right]
\end{align*}
where $\Lambda$ is the cosmological constant, the last four terms are the massive potential associate with graviton mass, $c_{i}$ are constants characterizing the strength of a fixed symmetric tensor
called the reference metric $f$ and $U_{i}$ are symmetric polynomials of the eigenvalue of the $(n+ 2)\times(n + 2)$ matrix $K^{\mu}{}_{\nu}=\sqrt{g^{\mu\alpha}f_{\alpha\nu}}$. We consider the static black hole solution with the follwing spacetime metric and reference metric as
\begin{align*}
&ds^2=-f(r)dt^2+\frac{dr^2}{f(r)}+r^2h_{ij}dx^idx^j,\\
&f_{\mu\nu}=diag(0,0,c_0^2h_{ij}),
\end{align*}
where $c_{0}$ is a positive constant, and $h_{ij}dx^idx^j$ is the line element for an Einstein space with constant curvature $n(n-1)k$ with $k = 1, 0, -1$ corresponding to a spherical, Ricci flat, or hyperbolic topology of the
black hole horizon, respectively.

We firstly focus on a class of four dimensional dS black holes, which have no additional dynamical constant characterizing the massive terms in the thermodynamical phase space. To including the effect of dynamical constant characterizing the massive terms, we then study a class of four dimensional AdS black holes in massive gravity.

\subsection{4D dS black hole in massive gravity}

Including the Maxwell field in the spacetime, the general charged black holes are presented in \cite{Nieuwenhuizen:2011sq,Cai:2014znn} with the horizon function $f(r)$
\begin{align*}
  f(r)=&k+\frac{16\pi P}{n(n+1)}r^2-\frac{16\pi M}{nV_n r^{n-1}}+
\frac{(16\pi Q)^2}{2n(n-1)V_n^2 r^{2(n-1)}}
+\frac{c_0c_1m^2}{n}r^2+c_0^2c_2m^2\\
&+\frac{(n-1)c_0^3c_3m^2}{r}
+\frac{(n-1)(n-2)c_0^4c_4m^2}{r^2}
\end{align*}
where the pressure is $P=-\frac{1}{8\pi}\Lambda=-\frac{n(n+1)}{16\pi \ell^2}$
and cosmological constant is $\Lambda=\frac{n(n+1)}{2\ell^2}$. We begin with four dimensional dS uncharged black hole in massive gravity with a reduced parameters choice
$k=1,n=2 ,V_n=4\pi,c_1=1,c_2=\frac{3a^2-1}{m^2}, c_3=c_4=0$. Then the metric function is simplified as
\begin{align}
  f(r)=3a^2-\frac{2M}{r}-\frac{r^2}{\ell^2},
\end{align}
with three horizons by solving its roots
\begin{align}
&R_E=2a\ell\sin{\left[\frac{1}{3}\arcsin{\left[\frac{M}{a^3\ell}\right]}\right]}\nonumber\\
&R_C=2a\ell\sin{\left[\frac{1}{3}\arcsin{\left[\frac{M}{a^3\ell}\right]}+\frac{2\pi}{3}\right]}\nonumber\\
&R_V=2a\ell\sin{\left[\frac{1}{3}\arcsin{\left[\frac{M}{a^3\ell}\right]}-\frac{2\pi}{3}\right]}
\end{align}
which correspond to the event horizon, cosmological horizon and ``virtual" horizon,respectively.

Thermodynamics of black holes in massive gravity are studied in \cite{Cai:2014znn,Xu:2015rfa}, here we list the temperature, entropy and thermodynamics volume in each horizon
\begin{align*}
&T_{i}=\frac{1}{4\pi}f'(r_i)=\frac{3a^2}{4\pi r_i}-\frac{3r_i^2}{4\pi \ell^2r_i}\\
&S_{i}=\int_0^{r_i}T^{-1}\left(\frac{\partial{M}}{\partial{r}}\right)_{P}\mathrm{d}r=
\frac{V_n}{4}r_i^n=\pi r_i^2\\
&V_{i}=\left(\frac{\partial{M}}{\partial{P}}\right)_{S}=\frac{V_n}{n+1}r_i^{n+1}=\frac{4\pi}{3}r_i^3.
\end{align*}
One can check the Smarr relations of three horizons as follow
\begin{align*}
&M=2(T_ES_E-V_EP)\\
&M=2(-T_CS_C-V_CP)\\
&M=2(T_VS_V-V_VP),
\end{align*}
where it is necessary to include the cosmological constant as thermodynamical viable in the thermodynamical phase space, other than the additional dynamical constant $a$ characterizing the strength of massive terms $U_{2}$. It is easy to obtain the entropy relations of this dS black hole
\begin{align}
&S_E S_C S_V=\frac{9M^2\pi}{16P^2},\nonumber\\
&S_E+S_C+S_V=-\frac{9a^2}{4P},\nonumber\\
&S_ES_C+S_ES_V+S_CS_V=\frac{81a^4}{64P^2},
\end{align}
where entropy product is mass-dependent while others are constants-dependent after including the effect of ``virtual" horizon.
After taking the first order derivative of these entropy relations, one can derive the first law of thermodynamic
\begin{align}
&dM=T_EdS_E+V_EdP,\nonumber\\
&dM=-T_CdS_C+V_CdP,\nonumber\\
&dM=T_VdS_V+V_VdP
\end{align}
where the following relations are used in three horizons
\begin{align*}
  &T_{E}=\frac{8P^2}{9M\pi}(S_E-S_C)(S_E-S_V),\quad\,T_{C}=\frac{8P^2}{9M\pi}(S_E-S_C)(S_C-S_V),\\
  &T_{V}=\frac{8P^2}{9M\pi}(S_E-S_V)(S_C-S_V),\quad\,V_{i}=\frac{M}{P}-\frac{a^2S_i(9a^2+8PS_i)}{4MP\pi}.
\end{align*}
Follow the same procedure as the discussion for Schwarzschild-dS black hole \cite{Xu:2015eia}, we find the entropy bounds for all horizons
\begin{align}
  S_E\in[0,\ell^2]\times\,a^2, \quad S_C\in[\ell^2,3\ell^2]\times\,a^2, \quad S_N\in[3\ell^2,4\ell^2]\times\,a^2
\end{align}
which are all geometrical inequalities in asymptotically non-flat spacetime. Besides, the constant of theory $a$
enlarges these bounds.

\subsection{4D AdS black hole in massive gravity}

We study four dimensional AdS black hole in massive gravity \cite{Nieuwenhuizen:2011sq,Cai:2014znn} with the cosmological constant $\Lambda=-\frac{n(n+1)}{2\ell^2}$ and special choices
$c_1=-\frac{2b^2}{m^2},c_2=\frac{3a^2-1}{m^2},c_3=c_4=0,k=1,n=2$.
Then the horizon function is
\begin{align}
  f(r)=3a^2-\frac{2M}{r}-b^2r+\frac{r^2}{\ell^2}.
\end{align}
These above parameters are under some constraints
which could lead to three positive horizons $r_1<r_2<r_3$ (roots of $f(r)$), and we do not attempt to interpret these constraints here.

For this case, one can obtain Smarr relation after the scale analysis
\begin{align}
&M=2(T_1S_1-V_1P+\frac{bV^b_1}{4}),\nonumber\\
&M=2(-T_2S_2-V_2P+\frac{bV^b_2}{4}),\nonumber\\
&M=2(T_3S_3-V_3P+\frac{bV^b_3}{4})
\end{align}
where $V^b_i=\left(\frac{\partial M}{\partial b}\right)\big|_{S,P}=br_i^2$ is the conjugate thermodynamic quantity associated with the additional dynamical constant $b$ characterizing the strength of  massive terms $U_1$.
Note that it is necessary to include both the constants of theory $\Lambda$ and $b$ as thermodynamical viable in the thermodynamical phase space. Then we list the mass-dependent entropy relations and constants-dependent entropy sum of this AdS black hole
\begin{align}
&S_1S_2S_3=\frac{9M^2\pi}{16P^2},\nonumber\\
&S_1S_2+S_1S_3+S_2S_3=\frac{9(9a^4-4b^2M)}{64P^2},\nonumber\\
&S_1+S_2+S_3=\frac{9b^4}{64P^2\pi}-\frac{9a^2}{4P}
\end{align}
The first order derivative of these entropy relations lead to the first law of thermodynamic of three horizons
\begin{align}
dM=T_1dS_1+V_1dP-V^b_1db,\nonumber\\
dM=-T_2dS_2+V_2dP-V^b_2db,\nonumber\\
dM=T_3dS_3+V_3dP-V^b_3db
\end{align}
where the conjugate thermodynamic quantity $V^b_{i}=\frac{bS_i}{\pi}$, thermodynamics volume $V_i=\frac{M}{P}+\frac{b^2S_i}{2P\pi}-\frac{a^2S_i(9a^2+8P S_i)}{2P(2M\pi+b^2S_i)}$ and the following temperature relations are used
\begin{align*}
&T_{1}=\frac{16P^2}{9(2M\pi+b^2S_1)}(S_1-S_2)(S_1-S_3),\\
&T_{2}=\frac{16P^2}{9(2M\pi+b^2S_2)}(S_1-S_2)(S_2-S_3),\\
&T_{3}=\frac{16P^2}{9(2M\pi+b^2S_3)}(S_1-S_3)(S_2-S_3).
\end{align*}
Similarly to the case for dS massive black hole, we can obtain the entropy bounds for three physical horizons
\begin{align}
&S_1\in[0,\frac{b^4\ell^4}{3}-2a^2\ell^2],S_2\in[\frac{b^4\ell^4}{3}-2a^2\ell^2,\frac{2b^4\ell^4}{3}-4a^2\ell^2],
S_3\in[\frac{2b^4\ell^4}{3}-4a^2\ell^2,b^4\ell^4-6a^2\ell^2],
\end{align}
where all above geometrical inequalities (entropy bounds) are modified by both the constants of theory $a$ and
$b$, while there is only the effect of constant $b$ characterizing the strength of massive terms $U_1$ in the thermodynamic phase space.

\section{Entropy relations and bounds in Einstein-dilaton gravity}

In this section, we show an example of entropy relations and bounds in Einstein-dilaton gravity. The gravity with dilaton
field is resulted from the low energy limit of string theory, and it has important consequences on the causal structure and the thermodynamic properties of black holes. The Einstein-Maxwell-Dilaton action in $(n+1)$-dimensional spacetime is \cite{Chan:1995fr}
\begin{align*}
  S=\frac{1}{16\pi}\int \mathrm{d^{n+1}}\sqrt{-g}\left[R-\frac{4}{n-1}\left(\nabla \Phi\right)^2-V(\Phi)-e^{-4\alpha\Phi/(n-1)}F_{\mu\nu}F^{\mu\nu}\right]
\end{align*}
where $V(\phi)$ is the dilaton potential and $V(0)=\Lambda$ is the cosmological constant. The theory admits a general charged dilaton black hole \cite{Chan:1995fr,Ong:2012yf}
\begin{align*}
&ds^2=-f(r)dt^2+\frac{dr^2}{f(r)}+r^2R^2d\Omega_{k,n-1}^2\\
&R=e^{\frac{2\alpha\phi}{n-1}},\quad \phi=\frac{(n-1)\alpha}{2(1+\alpha^2)}\log{\frac{b}{r}}
\end{align*}
with the metric function
\begin{align*}
  f(r)=&-\frac{k(n-2)(\alpha^2+1)^2b^{-2\gamma}r^{2\gamma}}{(\alpha^2-1)(\alpha^2+n-2)}
-\frac{m}{r^{(n-1)(1-\gamma)-1}}
+\frac{2q^2(\alpha^2+1)^2b^{-2(n-2)\gamma}}{(n-1)(\alpha^2+n-2)}r^{2(n-2)(\gamma-1)}\\
&+\frac{2\Lambda(\alpha^2+1)^2b^{2\gamma}}{(n-1)(\alpha^2-n)}r^{2(1-\gamma)},
\end{align*}
where $\Lambda=-\frac{n(n-1)}{(2\ell^2)}$ is the cosmological constant in general dimensions.

The conserved charges of charged topological dilaton black holes \cite{Sheykhi:2007wg} are related to the parameters in the solutions as
\begin{align*}
  &M=\frac{b^{(n-1)\gamma}(n-1)\omega_{n-1}}{16\pi(\alpha^2+1)}m,\quad\,Q=\frac{\omega_{n-1}}{4\pi}q.
\end{align*}
Other thermodynamic quantities, temperature and entropy are
\begin{align*}
 T_{r_h}=\frac{f'(r_h)}{4\pi}, \quad S=\frac{b^{(n-1)\gamma}\omega_{n-1}r_h^{(n-1)(1-\gamma)}}{4}.
\end{align*}
After treating the cosmological constant as thermodynamic pressure,
\begin{align*}
  P=-\frac{\Lambda}{8\pi}=\frac{n(n-1)}{16\pi \ell^2},
\quad V=-\frac{(\alpha^2+1)b^{\gamma(n+1)}\omega_{n-1}}{\alpha^2-n}r_h^{n-\gamma(n-1)}
\end{align*}
thermodynamics of charged topological dilaton black holes in the extend phase space are also studied \cite{Zhao:2013oza}.

For simplicity, we focus on a reduced four dimensional uncharged black holes with special parameters choice
$n=3,Q=0,k=1,\omega_{n-1}=4\pi,\alpha=\sqrt{2},b=1$, in order to give a qualitative study of entropy relations and bounds. Then the metric function becomes
\begin{align}
 f(r)=-3r^{\frac43}+\frac{27}{\ell^2}r^{\frac{2}{3}}-mr^{\frac{1}{3}}
\end{align}
which is similar with the Schwarzschild-AdS black hole. We can get the event horizon, Cauchy horizon and the ``virtual" horizon
\begin{align*}
&r_E^{1/3}=\frac{2\sqrt{3}}{\ell}\sin\left(\frac13 \arcsin\left(\frac{\sqrt{3}\ell^3m}{54}\right)\right),\\
&r_C^{1/3}=\frac{2\sqrt{3}}{\ell}\sin\left(\frac13 \arcsin\left(\frac{\sqrt{3}\ell^3m}{54}\right)+\frac{2\pi}{3}\right),\\
&r_V^{1/3}=\frac{2\sqrt{3}}{\ell}\sin\left(\frac13 \arcsin\left(\frac{\sqrt{3}\ell^3m}{54}\right)-\frac{2\pi}{3}\right).
\end{align*}
Thermodynamic quantities reduce to
\begin{align*}
&T_{i}=\frac{m}{12\pi r_{i}^{2/3}}-\frac{r_{i}^{1/3}}{2\pi},\quad S_{i}=\pi r_{i}^{2/3}\\
&P=\frac{3}{8\ell^2\pi},\quad V_{i}=12\pi r_{i}^{1/3},\quad\,M=\frac{m}{6}=4\pi\left(\frac{9r_{i}^{1/3}}{8\ell^2\pi}-\frac{r_{i}}{8\pi}\right)
\end{align*}
Consider thermodynamic laws in the extend phase space, the scaling argument leads to the Smarr relation
\begin{align}
&M=\frac23(T_ES_E+V_EP),\nonumber\\
&M=\frac23(-T_CS_C+V_CP),\nonumber\\
&M=\frac23(T_VS_N+V_VP).
\end{align}
As no charge and rotation in the spectime, one can obtain the mass-dependent entropy product
\cite{Majeed:2015eia}
\begin{align}
  S_ES_CS_V=4M^2\pi^3
\end{align}
and mass-independent UER
\begin{align}
  &S_E+S_C+S_V=48P\pi^2,\nonumber\\
  &S_ES_C+S_ES_V+S_CS_V=576P^2\pi^4
\end{align}
One can also derive the first law of thermodynamics by taking the first order derivative of these entropy relations
\begin{align}
dM=T_EdS_E+V_EdP,\nonumber\\
dM=-T_CdS_C+V_CdP,\nonumber\\
dM=T_NdS_N+V_NdP
\end{align}
Finally, a similar calculation based on these entropy relations leads to entropy bounds of three horizons
\begin{align}
S_E\in[0,3\pi\,\ell^2],\quad S_C\in [3\pi\,\ell^2,9\pi\,\ell^2],\quad
S_N\in[9\pi\,\ell^2,12\pi\,\ell^2].
\end{align}
which are all geometrical inequalities, even the spacetime does
not behave as (A)dS spacetime for the presence of a non-trivial dilaton.

\section{Entropy relations and bounds in Horava-Lifshitz gravity}

In this section, we generalize entropy relations and bounds to black holes in Horava-Lifshitz (HL) gravity \cite{Horava:2008ih,Horava:2009if,Horava:2009uw}, which is a quantum gravity model and have attracted a lot of attention. The action for HL gravity is
\begin{align*}
  S=\int \mathrm{d^3}x\mathrm{d}t&\sqrt{g}N\bigg[\frac{2}{k^2}(K_{ij}K^{ij}-\lambda K^2) +\frac{k^2\mu^2(\Lambda_wR-3\Lambda_w^2)}{8(1-3\lambda)}
+\frac{k^2\mu^2(1-4\lambda)}{32(1-3\lambda)}R^2\\
&\qquad-\frac{k^2}{2w^4}(C_{ij}-\frac{\mu w^2}{2}R^{ij})+\mu^4 R\bigg]
\end{align*}
with the Cotton tensor $C^{ij}=\epsilon^{ikl}\nabla_k(R_i^j-\frac{1}{4}\epsilon^{ikj}
\partial_k R)$ and other constants:  speed of light$c=\frac{k^2\mu}{4}\sqrt{\frac{\Lambda_w}{1-3\lambda}}$, Newton's constant $G=\frac{k^2}{32\pi c}$  and cosmology constant $\Lambda=\frac{3}{2}\Lambda_w$, which are all resulted from comparing the action with that of general relativity. The static solution behaviors as \cite{Lu:2009em,Cai2009}
\begin{align*}
  &ds^2=-N^2(r)dt^2+\frac{dr^2}{g(r)}+r^2(d\theta^2+\sin^2{\theta}d\varphi^2),\\
  &N^2(r)=g=1-\sqrt{4M\omega r+\omega^2r^4}+\omega r^2,
\end{align*}
where $M$ is an integration constant related to mass and the limit $\Lambda_w\rightarrow0$ is taking. The event horizon and Cauchy horizon are
\begin{align*}
  r_E=M+\sqrt{M^2-\frac{1}{2\omega}}\quad\,r_C=M-\sqrt{M^2-\frac{1}{2\omega}}
\end{align*}
Thermodynamic quatities, including the surface gravity, temperature and entropy are  \cite{Myung:2009dc,Myung:2009va,Cai:2009qs}
\begin{align*}
&\kappa_{i}=\frac{\omega(r_i-M)}{1+\omega r_{i}^2},\quad\,
T_{i}=\frac{\kappa_{i}}{2\pi}=\frac{\omega(r_i-M)}{2\pi(1+\omega r_{i}^2)},\quad\,
S_{i}=\frac{A_{i}}{4}=\pi r_i^2,
\end{align*}
respectively.

We show the entropy product and entropy sum \cite{Pradhan:2015fha,Mandal:2015ozx}
\begin{align}
 S_ES_C=\frac{\pi^2}{4\omega^2},\quad\,S_E+S_C=\pi\left(4M^2-\frac{1}{\omega}\right).
\end{align}
As the spacetime is asymptotically flat, the entropy sum behaviors as mass-dependent, while entropy product is
mass-independent, even there is no charges and rotation in the spacetime. This is the first example that the universality of entropy product holds for multi-horizons black holes without charges and angular momenta.
Then one can obtain the entropy bounds of both horizons directly
\begin{align}
  S_E\in[\pi(2M^2-\frac{1}{2\omega}),\pi\left(4M^2-\frac{1}{\omega}\right)],\quad\,S_C\in[0,\frac{\pi}{2\omega}]
\end{align}
with the area bounds
\begin{align}
  \sqrt{\frac{A_E}{16\pi}}\in\left[\frac{M}{\sqrt{2}},
M\right]\times\sqrt{1-\frac{1}{4M^2\omega}},\quad\,\sqrt{\frac{A_C}{16\pi}}\in\left[0,\frac{1}{2\sqrt{2\omega}}\right].
\end{align}
Especially for the upper bound of area for event horizon, it is Penrose inequality of black hole \cite{Mars:2009cj} modified slightly by the constant of theory, while it is actually the exact Penrose inequality when $\omega\rightarrow0$ that the theory reduces to general relativity.

\section{Discussion}

In this paper, we review the so-called universal area (entropy) product formulae and its generalizations and extensions. We summarize and expand recent developments on the UER of black holes with multi-horizons, even including the effect of ``virtual'' horizon.
We show the universal entropy product which does not depend on the mass, together with some failed examples. Furthermore, we present others UER including the entropy sum. The universality of these area (entropy) relations contains quantization features that they only depend on the quantized charges (including
the electric charges and angular momenta) and the cosmological constant. Entropy product may be mass-independent for charged rotating black holes, while the universality of entropy sum holds in asymptotically non-flat spacetime.

Based on these entropy relations, one can obtain some entropy bounds, which are consistent with the Penrose inequality. Especially for the AdS spacetime, one should consider the cosmological constant
as thermodynamic pressure, hence we obtain the entropy bound which is actually a geometrical inequality. We expand the discussion about entropy relations in massive gravity, Einstein-dilaton gravity and Horava-Lifshitz gravity, where the constants of theory should also be treated as thermodynamic variables.  We obtain new entropy bounds and thermodynamics laws in these modified gravity models.

However, UER is almost an phenomenological approach, holding quantum features, to describe the microscopic entropy.
In this sense, it is more important to calculating the black hole entropy from these UER. Especially in the side of conformal field theory, one can find UER depend on the number of left/right-moving models, and vice versa. When one find the UER for general black holes, one can built the left/right-moving models in someway, and hence lead to the physical picture for the statistical origin of black holes. The black hole/CFT (BH/CFT) correspondence \cite{Chen:2012mh,Chen:2012yd,Chen:2012ps,Chen:2012pt,Chen:2013rb,Chen:2013aza,Chen:2013qza} open a window for this viewpoint, where the mass-independent entropy product is useful to built the
holographic descriptions of black hole entropy, for black hole with two horizons,.

Another important subject of microscopic black hole entropy is the physical meaning of microstate of black holes entropy\footnote{one can see a recent interesting idea in \cite{Wei:2015iwa}, where the number density of the black hole molecules is introduced to investigate the possible  microscopic structure of a charged anti-de Sitter black hole completely from the thermodynamic viewpoint.}. Because of the close relationship of multi-horizons in UER, one should consider the physical meaning of microstate of all horizons together in general background. Besides, it is interesting to study thermodynamics in Cauchy horizon, especially for the stability and phase transition.


\acknowledgments{Acknowledgments}
Wei Xu was supported by the Natural Science Foundation of China (NSFC) under Grant No.11505065 and the China Postdoctoral  Science Foundation under Grant No.2015M572130. This work is partially supported by the Natural Science Foundation of China (NSFC) under Grant No.11075078.





\conflictofinterests{Conflicts of Interest}

The authors declare no conflict of interest.

\bibliographystyle{mdpi}
\makeatletter
\renewcommand\@biblabel[1]{#1. }
\makeatother

\providecommand{\href}[2]{#2}\begingroup
\footnotesize\itemsep=0pt
\providecommand{\eprint}[2][]{\href{http://arxiv.org/abs/#2}{arXiv:#2}}



%


%

\end{document}